\documentclass{article}  %>>> use for US letter paper
\usepackage{amsmath,amsfonts,amssymb}
\usepackage{graphicx}
\usepackage{fancyhdr}
\usepackage{rotating}
\usepackage{color}

\newcommand{\be}{\begin{equation}}
\newcommand{\ee}{\end{equation}}

\title{The power spectrum extended technique applied to images of binary stars in the infrared} 

\author{Eric Aristidi$^a$, Eric Cottalorda$^a,b$, Marcel Carbillet$^a$, Lyu Abe$^a$, \\
Karim Makki$^c$, Jean-Pierre Rivet$^a$, David Vernet$^d$, Philippe Bendjoya$^a$}
\date{\sl\small
$^a$Universit\'e C\^ote d'Azur, Observatoire de la C\^ote d'Azur, CNRS, laboratoire Lagrange, France,
\\$^b$ArianeGroup, 51/61 route de Verneuil - BP 71040, 78131 Les Mureaux Cedex, France\\
$^c$Laboratoire d'informatique et syst\`emes, Aix-Marseille Universit\'e, France\\
$^d$Universit\'e C\^ote d'Azur, Observatoire de la C\^ote d'Azur, France}

\begin{document} 
  %\bigskip
  
  \maketitle 

%%%%%%%%%%%%%%%%%%%%%%%%%%%%%%%%%%%%%%%%%%%%%%%%%%%%%%%%%%%%% 
\section*{Abstract}
We recently proposed a new lucky imaging technique, the Power Spectrum Extended (PSE), adapted for image reconstruction of short-exposure astronomical images in case of weak turbulence or partial adaptive optics correction. In this communication we show applications of this technique to observations of about 30 binary stars in H band with the 1m telescope of the Calern C2PU observatory. We show some images reconstructed at the diffraction limit of the telescope and provide measurements of relative astrometry and photometry of observed couples.

%%%%%%%%%%%%%%%%%%%%%%%%%%%%%%%%%%%%%%%%%%%%%%%%%%%%%%%%%%%%%
\section{INTRODUCTION}

\label{par:intro}  % \label{} allows reference to this section
The speckle interferometry technique introduced by Labeyrie\cite{Labeyrie70} is well adapted to the measurement of binary stars relative astrometry (and sometimes differential photometry), and active groups are still using it routinely to monitor double stars motions and refine orbits\cite{Mason18,Scardia20}. Its major drawback is that it cannot provide true images, and several improvements were proposed, generally based on the computation of higher-order statistical quantities of the speckle patterns. The bispectral analysis\cite{Weigelt91} is generally recognized as state-of-the art of the speckle imaging, but it is a somewhat heavy process that requires a lot of CPU time and memory.

The Lucky imaging (LI)\cite{Fried78} is a simple alternative for modest telescopes or in case of weak turbulence: Huffnagel\cite{Huffnagel66} calculated that LI techniques give good performances when $D/r_0\lesssim 7$, $D$ being the telescope diameter and $r_0$ the Fried parameter. LI processing have be combined successively with Adaptive Optics\cite{Velasco18}, making is efficient with larger telescopes or poorer seeing conditions. The LI technique relies on two essential points: the image selection and the alignment process. A review was made by Garrel et al.\cite{Garrel12} who proposed a selection criterion based on the Fourier transform of images; their technique, referred to as Fourier-Lucky (FL) imaging hereafter, is today one of the most efficient LI algorithms.

We recently began to work on a new technique\cite{Cottalorda19}, the ``Power Spectrum-Extented'', which is a combination of a LI and speckle interferometry. The algorithm is described in details in this conference by Cottalorda et al.\cite{Cottalorda20}. It was developed within the framework of the Calern Imaging Adaptive Optics (CIAO) project\cite{Carbillet20}. This AO bench is developed at the Epsilon telescope (diameter 1.04m) of the Centre P\'edagogique Plan\`etes et Univers (C2PU) facility\cite{Bendjoya12} located at the Calern observing site (South of France), and is designed to operate in the visible and the near-infrared (H-band). Several test campaigns were made during the conception of this instrument; one of them was dedicated to binary stars observation in the infrared: it is the purpose of this presentation.

%%%%%%%%%%%%%%%%%%%%%%%%%%%%%%%%%%%%%%%%%%%%%%%%%%%%%%%%%%%%%
\section{OBSERVATIONS AND DATA PROCESSING}
\begin{figure}
\includegraphics[width=8cm]{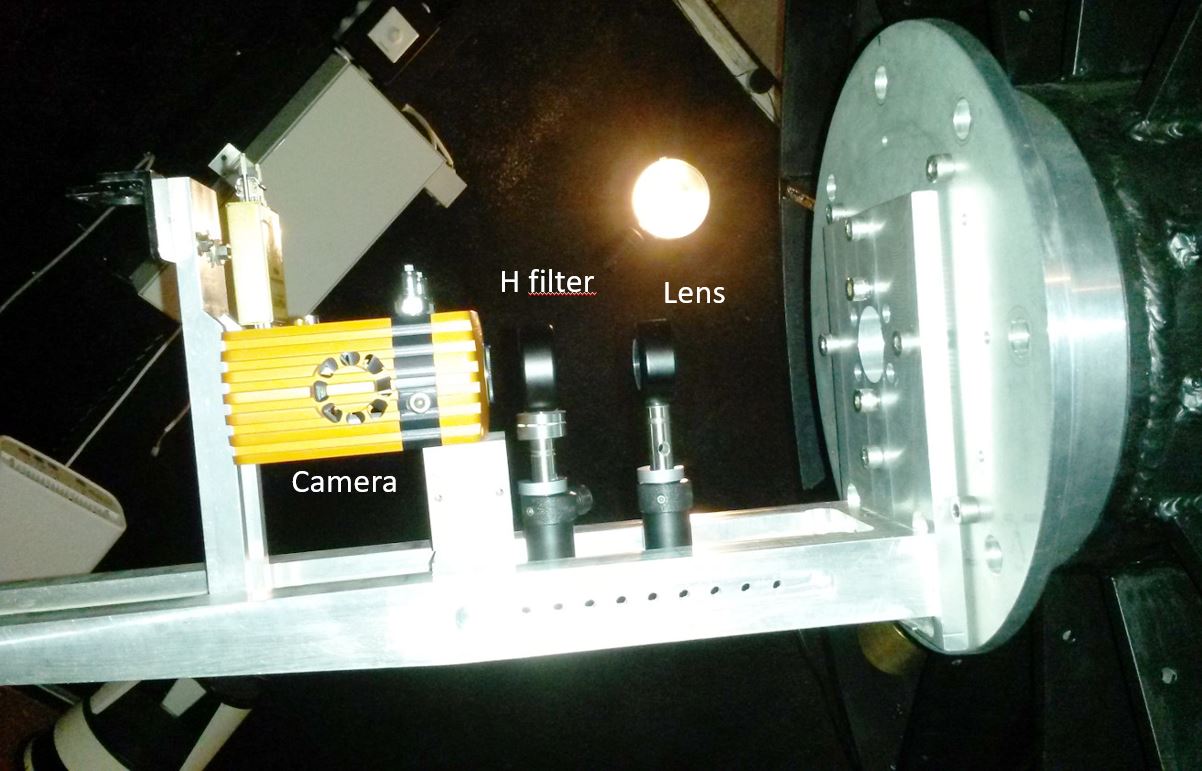}\hskip 1cm
\includegraphics[width=7cm]{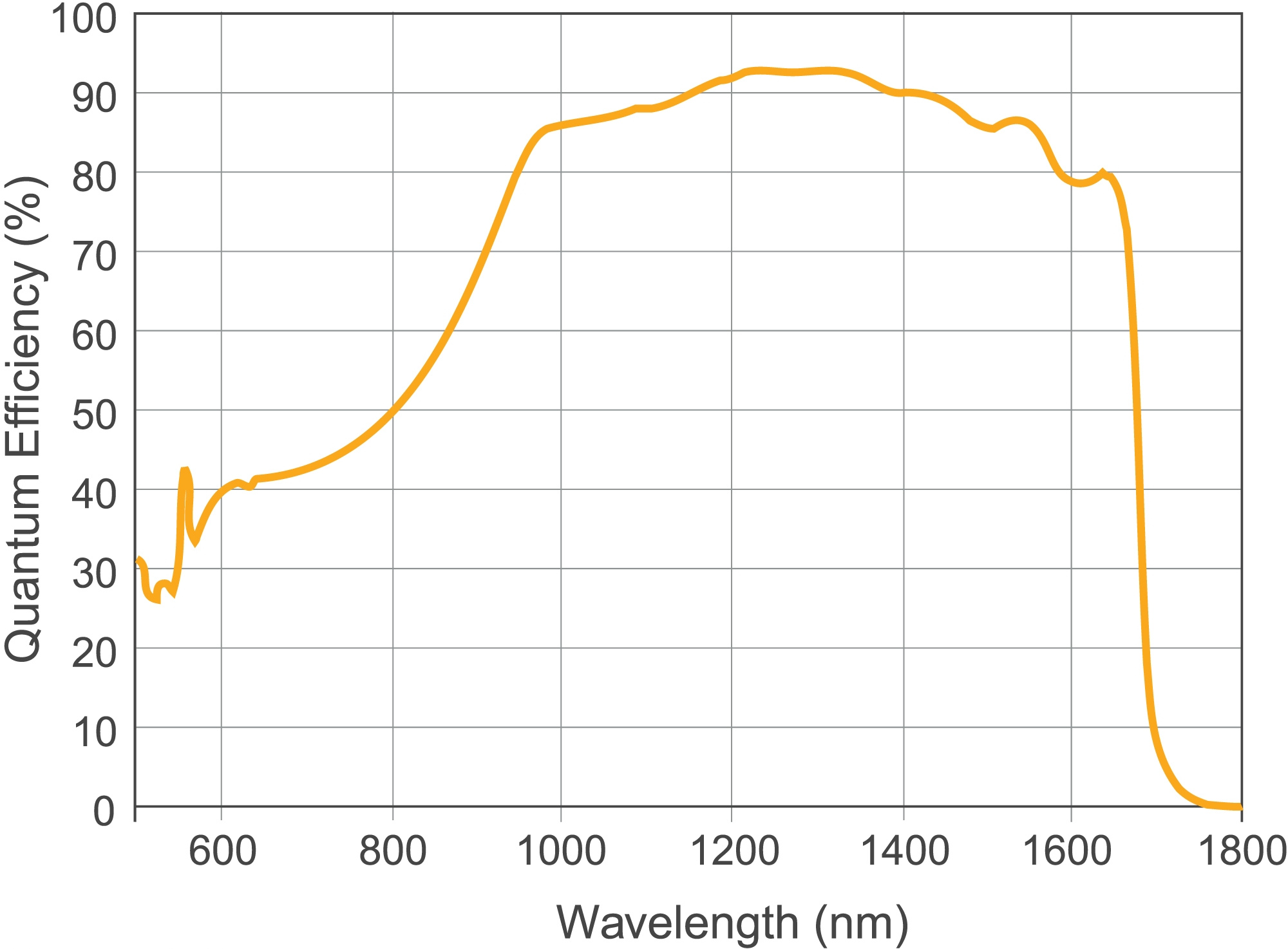}
\caption{Left: Optical bench at the Cassegrain focus. Right: Quantum efficiency curve of the Ninox 640 sensor (from Raptor website).}
\label{fig:benchQEcurve}
\end{figure}

Observations were carried out with the Epsilon telescope. It was used in Cassegrain configuration (focal ratio of 12.5), equipped with a simple optical bench hosting an infrared H filter (central wavelength $\lambda=1650$~nm, bandpass $\delta\lambda=350$~nm), a magnification lens and a near-infrared camera Ninox SWIR 640 from Raptor Photonics\cite{ninox}. A photo of the bench is shown in Fig.~\ref{fig:benchQEcurve}

The main characteristics of the camera are displayed in Table~\ref{table:camera}. It proposes two gain modes (low-gain and high-gain), all our observations were made with the high-gain mode. The camera allows short-exposure time down to 10ms in high gain mode, and is well adapted to speckle or lucky imaging applications. It is cooled down to -20$^\circ$C by a Peltier module, even more with the circulation of a cooling liquid (that we did not use for the present observations).
 
However it has a somewhat strong noise which limit performances. Hence the limiting magnitude of our instrumentation is about $H=6$ with an exposure time of 10~ms. The angular resolution of the telescope at $\lambda=1.6\mu$m is 0.3 arcsec. A few hundreds of bright double stars with an angular separation $\rho>0.3''$ and a magnitude $H<6$ were selected for observations from the Washington Double Star (WDS) catalog\cite{wdscatalog}. Observations were carried out in June 2016. About 60 binary stars were observed during 8 nights, leading to 27 positive detections of companions which are presented in Table~\ref{table:mesur}. Negative detections can be due to a too large magnitude difference, or bad seeing conditions.

The median seeing at Calern in the visible ($\lambda=500$nm) is 1.1$''$\cite{Aristidi19}, corresponding to 0.9$''$ in the H band ($\lambda=1650$nm). Indeed during our observing run, the seeing monitor recorded values between 0.6$''$ and 3$''$ at $\lambda=500$nm (0.5$''$ and 2.4$''$ at $\lambda=1650$nm). This corresponds to weak $D/r_0$ ratios (between 1.5 and 7) which are totally adapted to PSE applications. In particular the night of June 23rd to 24th, conditions were exceptional with a seeing (at $\lambda=500$nm) between 0.6 and 0.8~arcsec ($D/r_0$ between 1.7 and 1.9  in H~band) during the full night, and we obtained almost diffraction-limited images.

The coherence time was very low during the first nights (around 2ms at $\lambda=1650$nm), and increased up to 80ms (at $\lambda=1650$nm) during the night of June 23rd to 24th (indeed the coherence time is proportional to the Fried parameter\cite{Aristidi19} and good seeing conditions correspond generally to slow turbulence). That night, we could increase the exposure time to 100ms to gain sensitivity.

\subsubsection*{Data processing}
\label{par:dataproc}
For each object a cube of several thousands of short-exposure images was recorded. After subtraction of the mean dark current (mandatory for this camera), image cubes were processed using the PSE algorithm\cite{Cottalorda20}. We recall here the main steps:
\begin{itemize}
\item Classify image quality according to a criterion based on the estimation of an instantaneous Fried parameter $r_0$, and select a percentage of images for processing,
\item Align all selected images to the first one (or to the best one). This alignment is performed in the Fourier plane, making use of the phase of the cross-spectrum between images,
\item Sum centered images, and extract the phase of the Fourier transform of this sum,
\item Calculate the average power spectrum of images, and extract the modulus of the Fourier transform of the object (following the well known speckle interferometry technique introduced by Labeyrie\cite{Labeyrie70}),
\item From the modulus and the phase of the Fourier transform calculated above, derive the final image.
\end{itemize}
This technique appeared to be fast and efficient on our images. Some comparisons with other data processing methods (shift-and-add (SA), FL, speckle) are presented in Sect.~\ref{par:results}. Accurate relative astrometry and photometry could be extracted from reconstructed images of the binary stars, even in the worst seeing conditions.

\subsubsection*{Scale calibration}
The scale and position angle calibration was done by taking a sequence of short-exposure images of the bright and large double star $\zeta$Uma (STF~1744AB). This slow motion couple does not have an orbit yet: its position angle has moved by only 10$^\circ$ since its first measurement in 1755\cite{wdscatalog}, and it is therefore a good object for calibration. A recent measurement\cite{Harshaw15} (epoch 2015.351) gave a separation of $\rho=14.45\pm 0.05''$ and a position angle $\theta=153.0\pm 0.3^\circ$. To calculate the pixel scale, we computed the average autocorrelation of images of the binary (this gives a better accuracy than a classical shift-and-add algorithm). This kind of processing is well known in speckle interferometry to measure double-star separation. This function exhibits 3 peaks whose distance is the separation of the binary stars in pixels. This gave a pixel scale of $\xi=0.0782\pm 0.0003$~arcsec. Uncertainties on the pixel scale and the position angle of this calibration star are taken in account in the error bars on $\rho$ and $\theta$ presented in the table~\ref{table:mesur}.

\begin{table}
\begin{tabular}{ccccccccc}\hline
Nb of   &   Pixel        &  Dynamic     &  Expos.  &  Frame       & Bandwidth &  RON & Dark & Cooling\\
pixels  &   size         &   range      & time     &  rate       &  ($\mu$m)   &      &  current & temp.\\ \hline
640$\times$512 & 15$\times$15 $\mu$m  & 14 bits     &  10ms--27s     & $\le$120 Hz &  0.4--1.7     &  37e$^-$ & 1500 e$^-$/s  & -20$^\circ$C\\ \hline
\end{tabular}
\caption{Characteristics of the Ninox 640 camera (High-Gain mode)}.
\label{table:camera}
\end{table}

%\begin{table}
%\begin{tabular}{l|l}
%Nb of   pixels &  640$\times$512  \\
%Pixel  size  &   15 $\mu$m   \\
%Dynamic range &  14 bits  \\
%Expos. time  &  \\
%Frame rate & 10--120 Hz  \\
%Bandwidth &  0.4--1.7 $\mu$m \\
%RON &  18e$^-$  \\
%Cooling temp. & -20$^\circ$C  \\
%\end{tabular}
%\caption{Characteristics of the Ninox 640 camera}.
%\label{table:camera}
%\end{table}

%%%%%%%%%%%%%%%%%%%%%%%%%%%%%%%%%%%%%%%%%%%%%%%%%%%%%%%%%%%%%
\begin{figure}
\includegraphics[width=65mm]{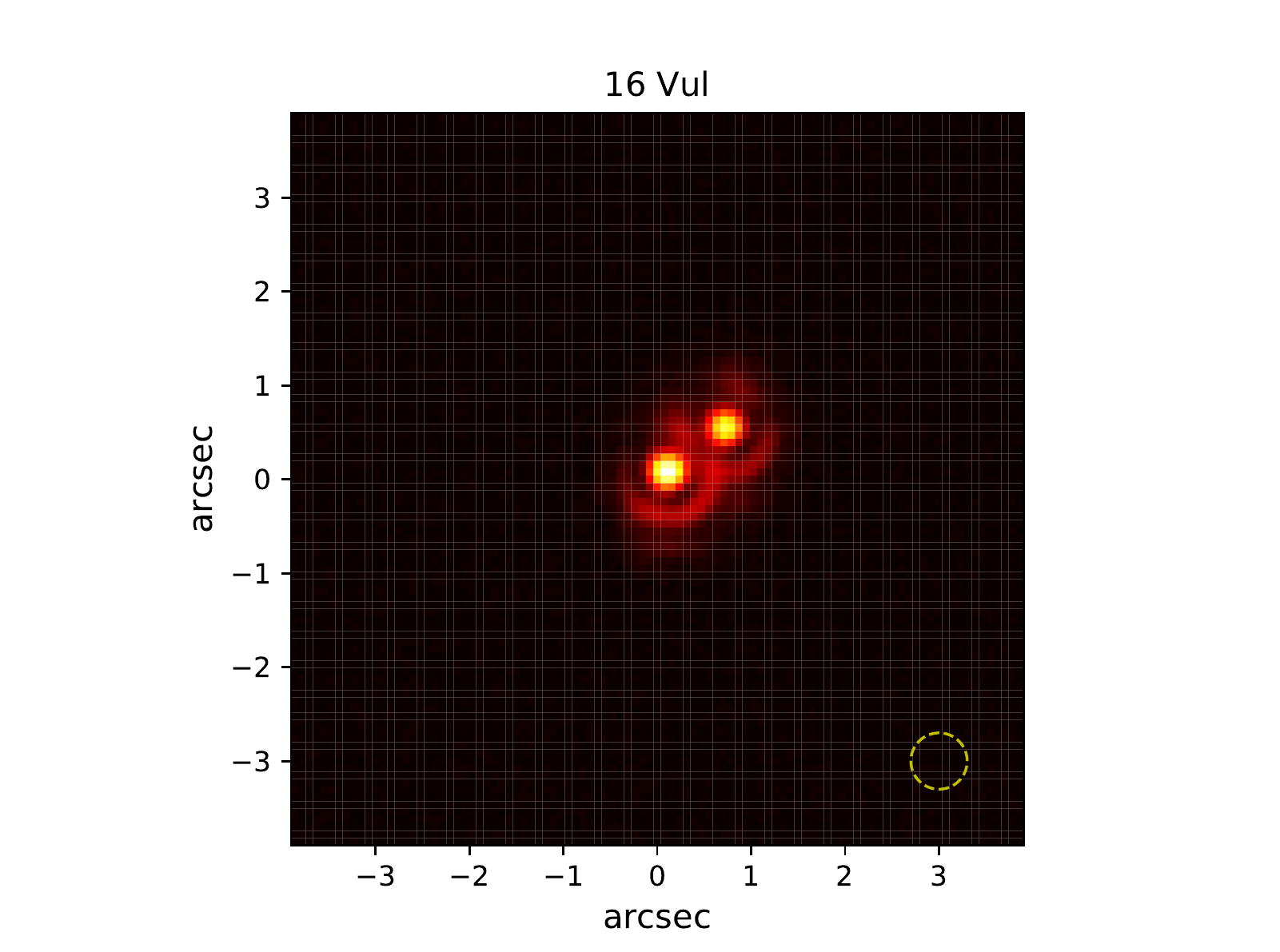} \hskip -10mm
\includegraphics[width=65mm]{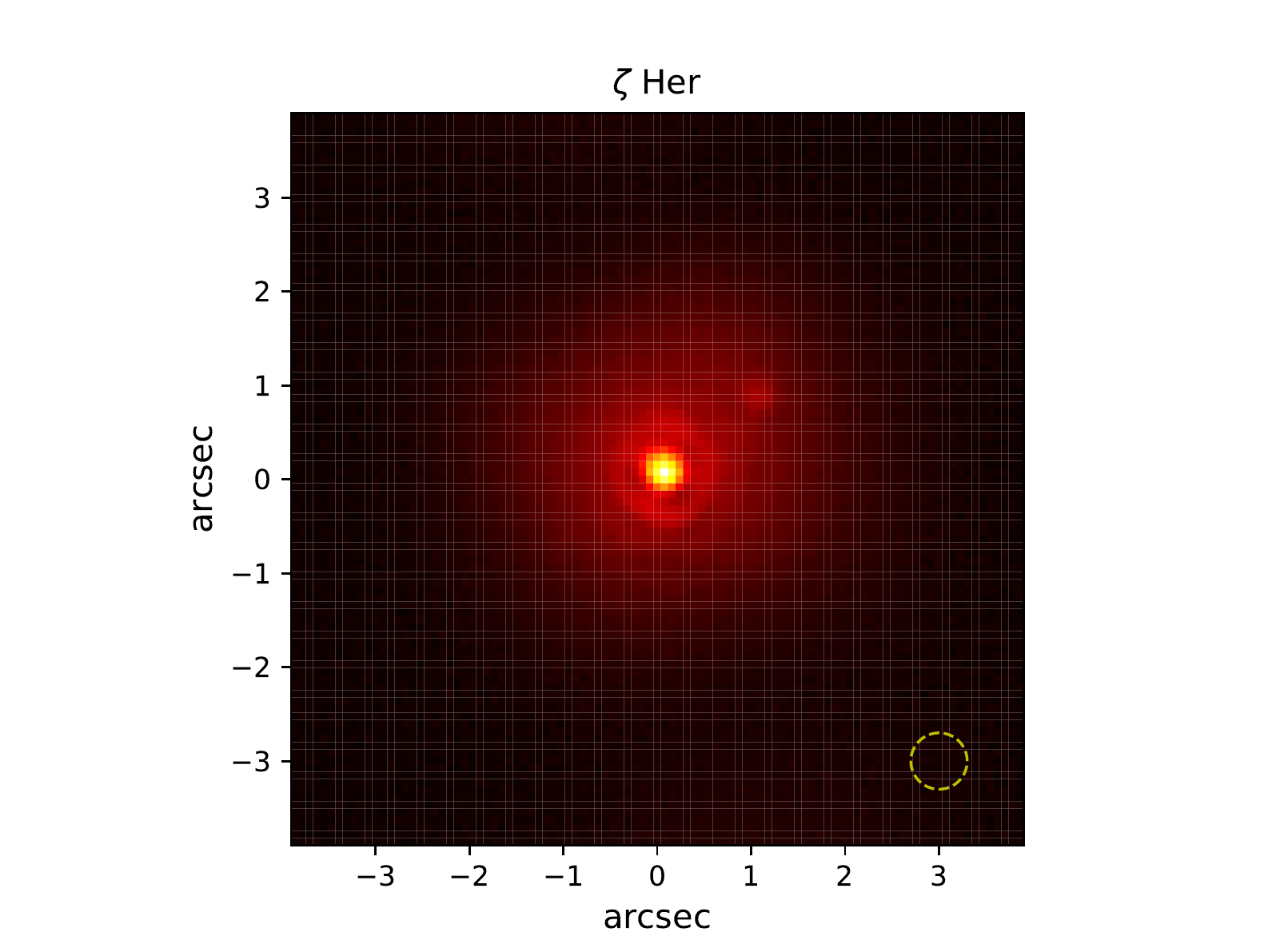} \hskip -10mm
\includegraphics[width=65mm]{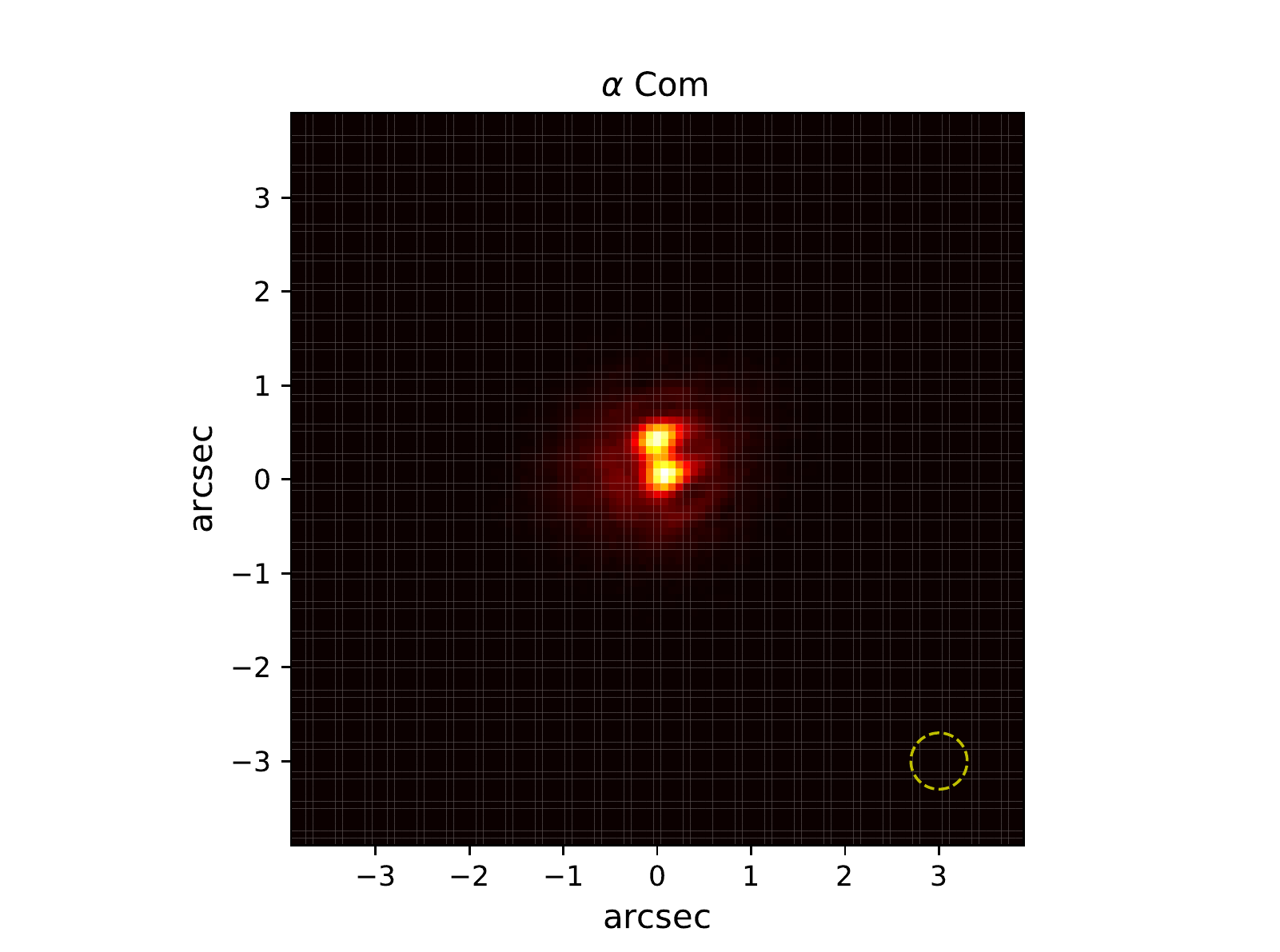} \\
\caption{Exemple of images reconstructed by the PSE algorithm, for the 3 stars 16Vul ($\rho=0.75''$), $\zeta$Her ($\rho=1.27''$) and $\alpha$Com ($\rho=0.36''$) The yellow circle at the bottom right corner represents the Airy disc size (radius $\lambda/D=0.3''$).}
\label{fig:expse}
% script python : img_4stars.py
\end{figure}

\section{RESULTS}
\label{par:results}
Fig.~\ref{fig:expse} shows examples of reconstructed images by the PSE algorithm, for 3 binary stars of different separations. Images appear to be diffraction-limited and exhibit Airy rings around stars. In particular, the pair $\alpha$Com is well resolved, while it is close to the diffraction limit of the telescope ($\rho=0.36''$). 

\begin{figure}
\includegraphics[width=65mm]{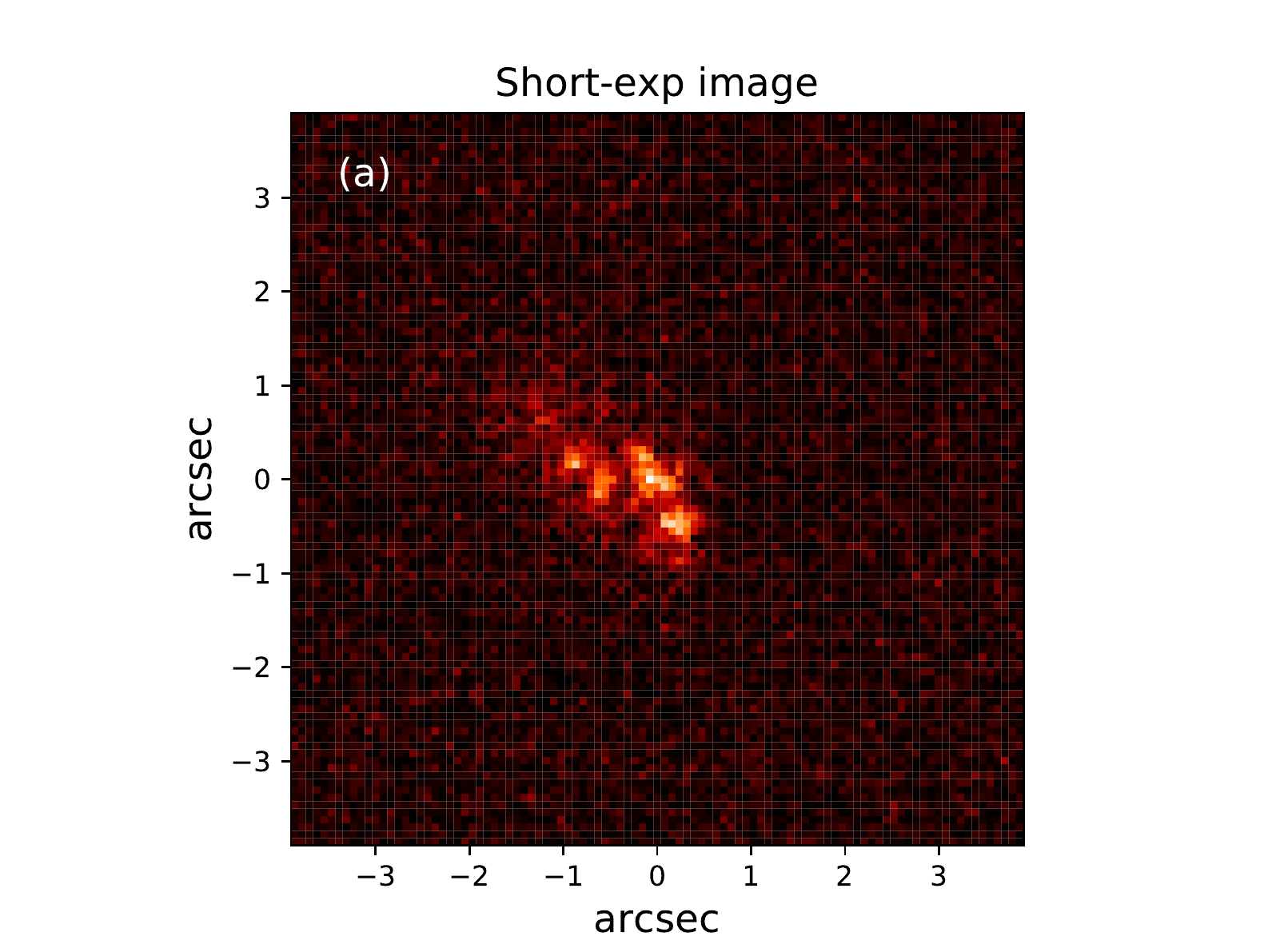} \hskip -10mm
\includegraphics[width=65mm]{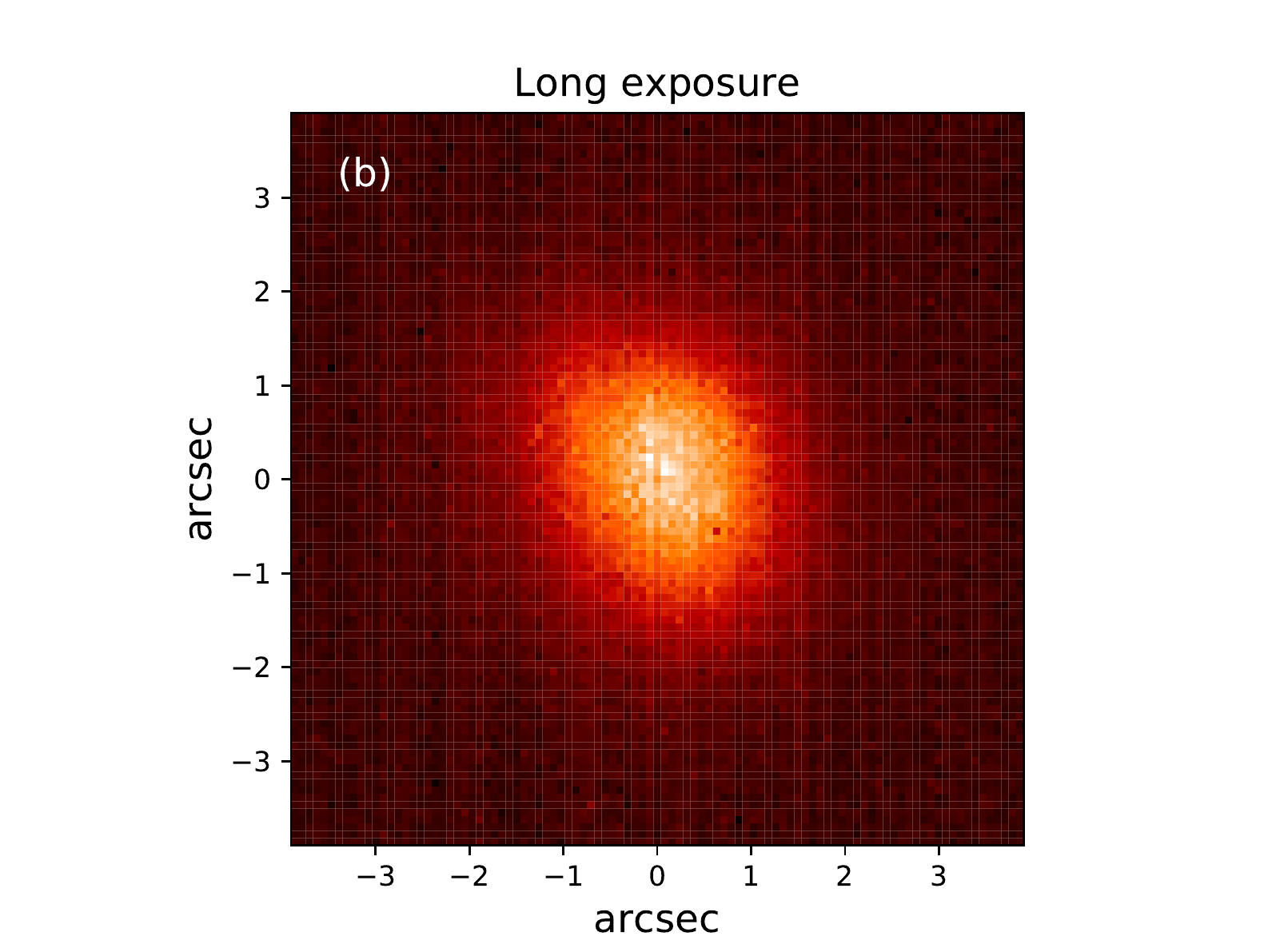} \hskip -10mm
\includegraphics[width=65mm]{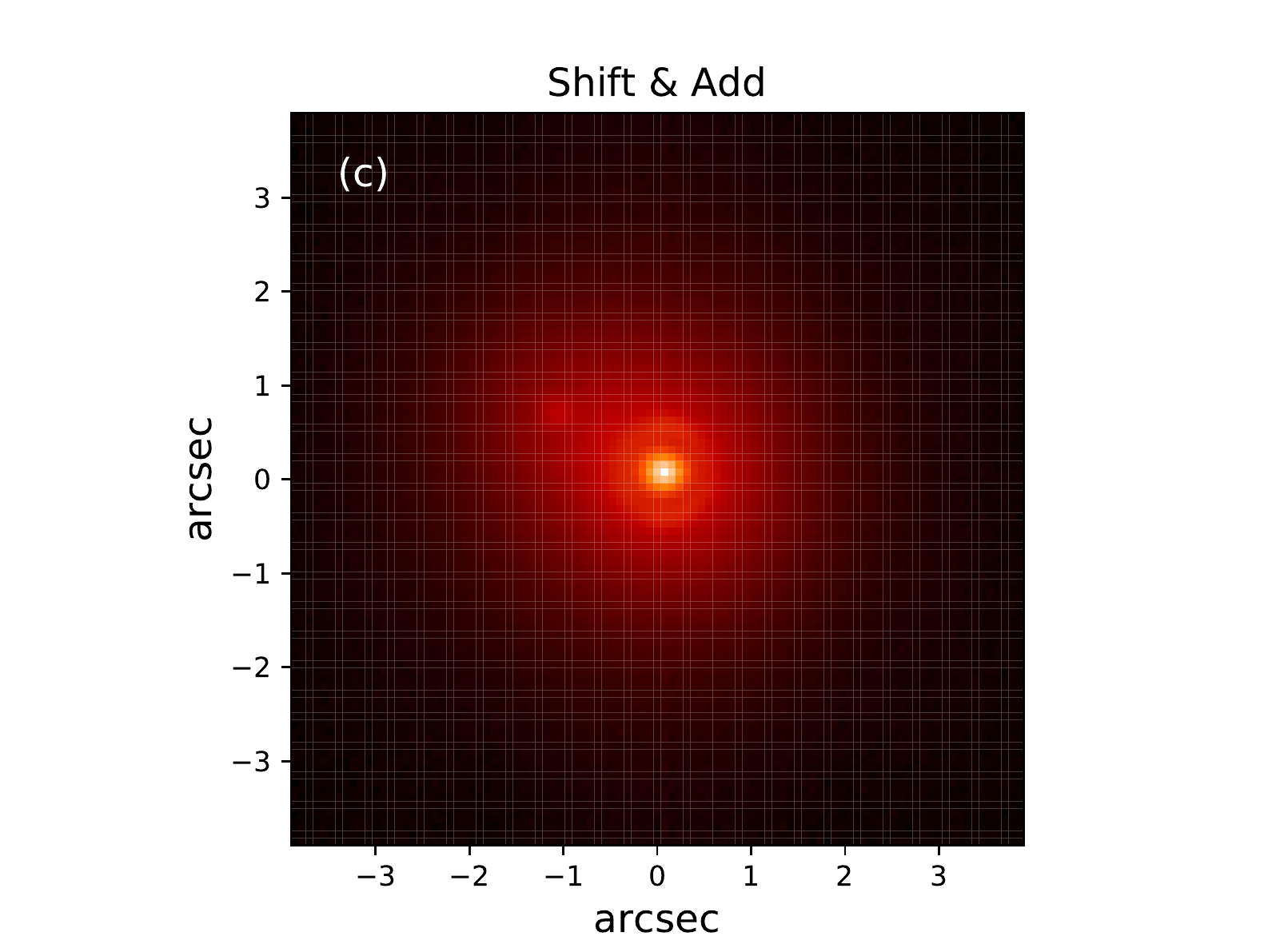} \\
\includegraphics[width=65mm]{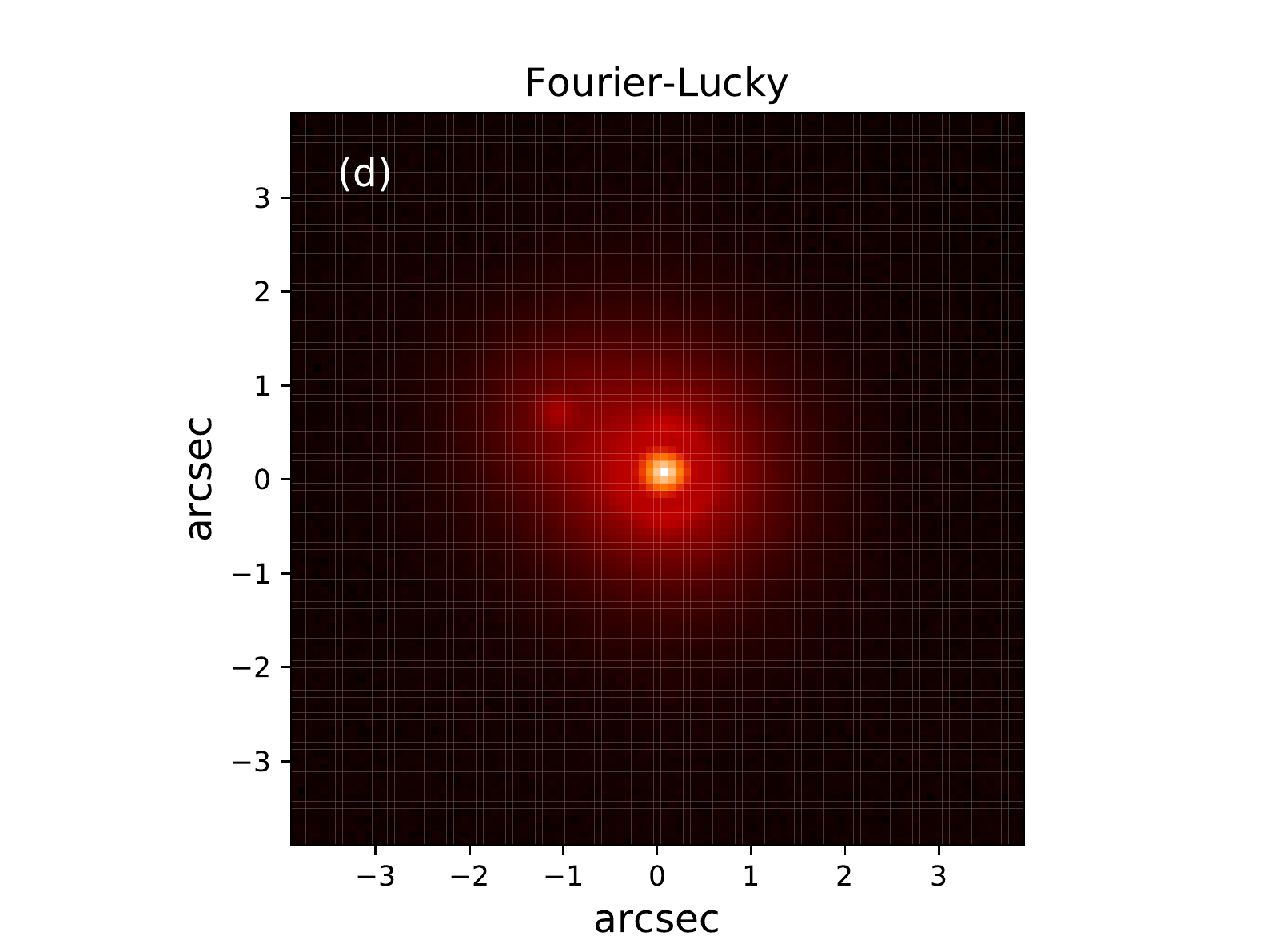} \hskip -10mm
\includegraphics[width=65mm]{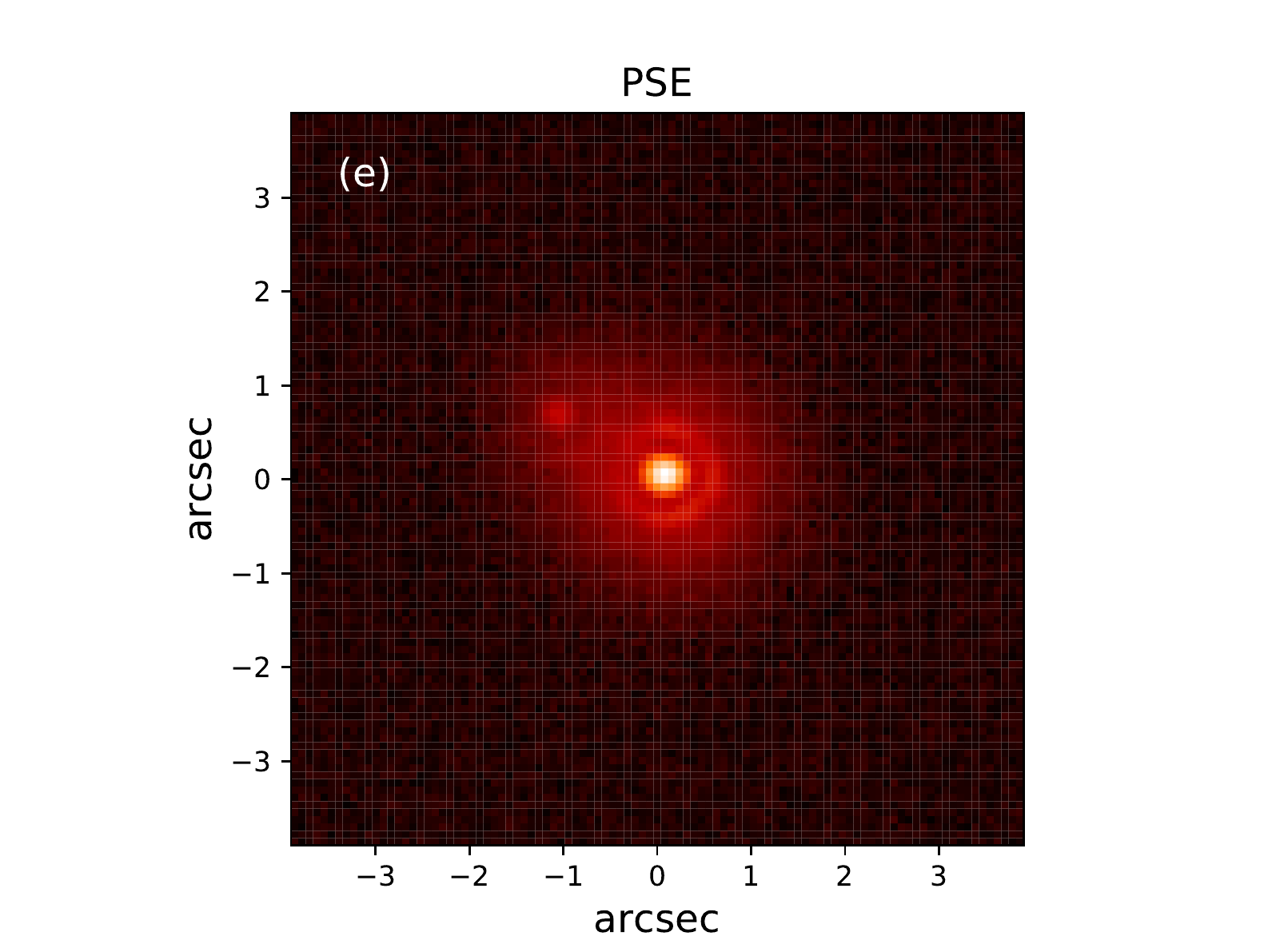} \hskip -10mm
\includegraphics[width=65mm]{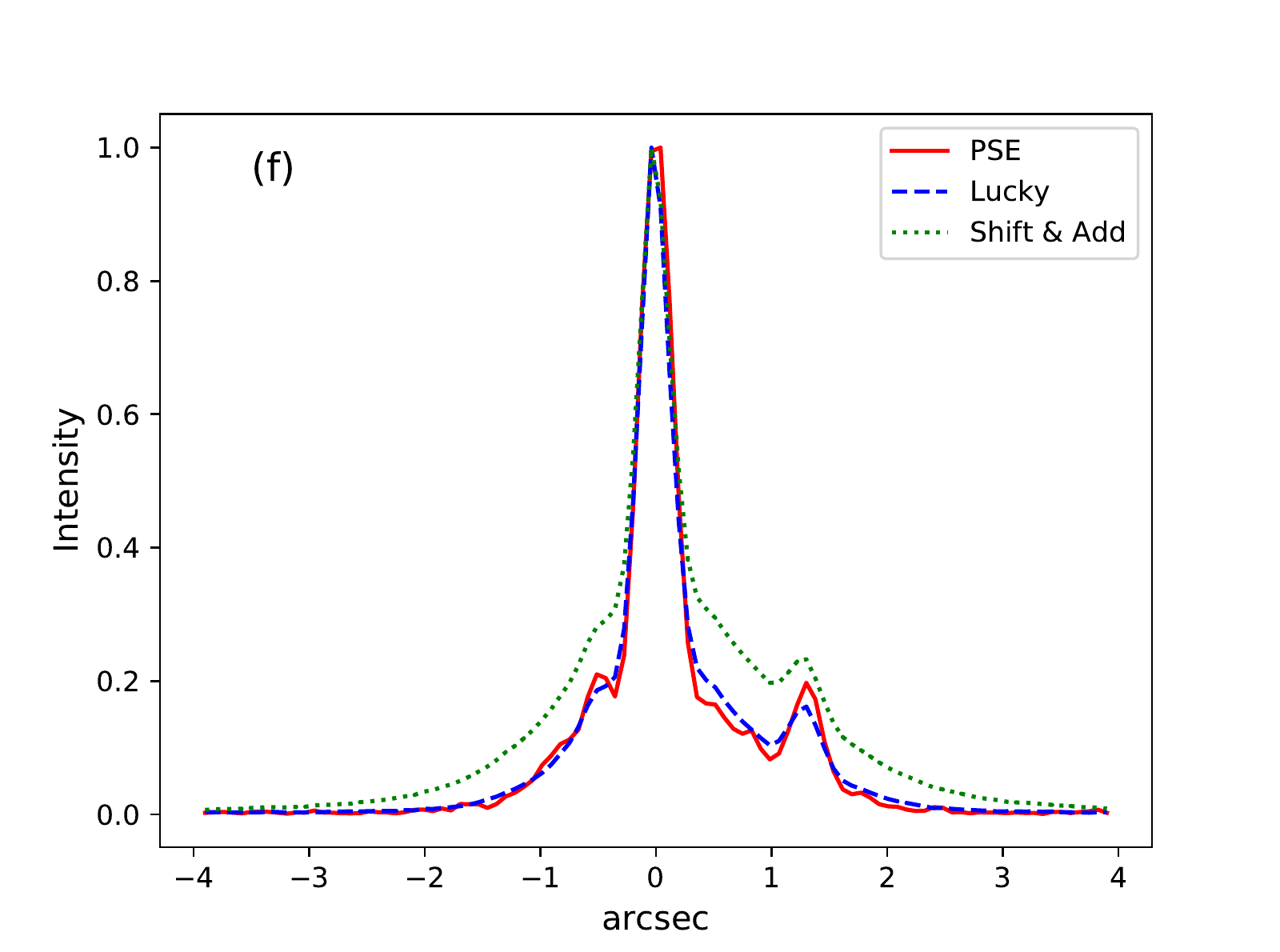}
\caption{Data processing on the star ADS 11871. (a) example of a raw image. (b) long-exposure image. (c) Reconstructed image by SA.  (d) Reconstructed image by FL with 50\% selection. (e) Reconstructed image by PSE processing (50\% selection). (f) Intensity plot of reconstructed images along the line joining the stars.}
\label{fig:compar}
% script python : plot_coup_img_compar.py
\end{figure}

A comparison between different data processing techniques is shown in Fig.~\ref{fig:compar}. We took the example of the double star ADS~11871 which appear to be a good test case since it has a large magnitude difference ($\Delta m=1.8$) and it was observed under bad seeing conditions ($\epsilon=2.5''$ at $\lambda=500$nm corresponding to $2''$ at $\lambda=1650$nm). The long-exposure image (b) was obtained by adding the whole set of 13\,000 individual frames. The SA image (c) was computed by centering all frames on the brightest pixel, them adding the whole image set. It is the fastest and the easyest of the algorithms that we tested, and it clearly shows the binary. However the image is embedded is a diffuse halo of the size of the seeing disc. The FL image (d) was made using Fourier-lucky algorithm\cite{Garrel12}. The centering is made, as for the SA, on the brightest pixel. The image shown was calculated from a selection of the 50\% best frames (according to the criterion described in Sect.\ref{par:dataproc}). As expected, there is a substantial improvement over the SA image. The PSE image (e) was also computed on a selection of the best 50\%, and its quality is close to the FL image, at one can see on profiles displayed on the plot (f). We could estimate the Strehl ratio $S_t$ of reconstructed images, using the formula by Tokovinin\cite{Tokovinin02} after having removed the companion star from images. We found
\begin{itemize}
\item Image (c), SA : $S_t=8$\%
\item Image (d), FL : $S_t=20$\%
\item Image (e), PSE : $S_t=24$\%
\end{itemize}
showing a slight advantage of the PSE method on the FL one, for this particular example. Table~\ref{table:compar} gives the average error bars on mesurements of the separation $\rho$ and the position angle $\theta$ for the four processing methods (SA, FL, speckle and PSE). Speckle remains the most accurate technique, but we can see that PSE and FL give comparable uncertainties and are far better than SA. Same for the average Strehl ratios dispayed for the 3 techniques SA, FL, PSE in Table~\ref{table:compar}.

\begin{table}
\begin{center}
\begin{tabular}{cccc|cccc|ccc}\hline
\multicolumn{4}{c|}{$\delta\rho$ (arcsec)} & \multicolumn{4}{c|}{$\delta\theta\; (^\circ)$} & \multicolumn{3}{c}{Strehl ratio (\%)}\\
SA & FL & Speckle & PSE &       SA & FL & Speckle & PSE   & SA   &  FL  & PSE\\ \hline
0.16 & 0.13 & 0.11 & 0.14 &    0.79 & 0.54 & 0.50 & 0.52 &  15 & 41 & 36 \\ \hline
\end{tabular}
\end{center}
\caption{Average uncertainty on the separation ($\delta\rho$), position angle ($\delta\theta$) measurements ans Strehl ratios for the different processing techniques. No Strehl ratio is given for speckle since it is a measurement on autocorrelations.}
\label{table:compar}
\end{table}

\subsubsection*{The ``ghost'' effect}
\begin{figure}
\begin{center}
\includegraphics[width=65mm]{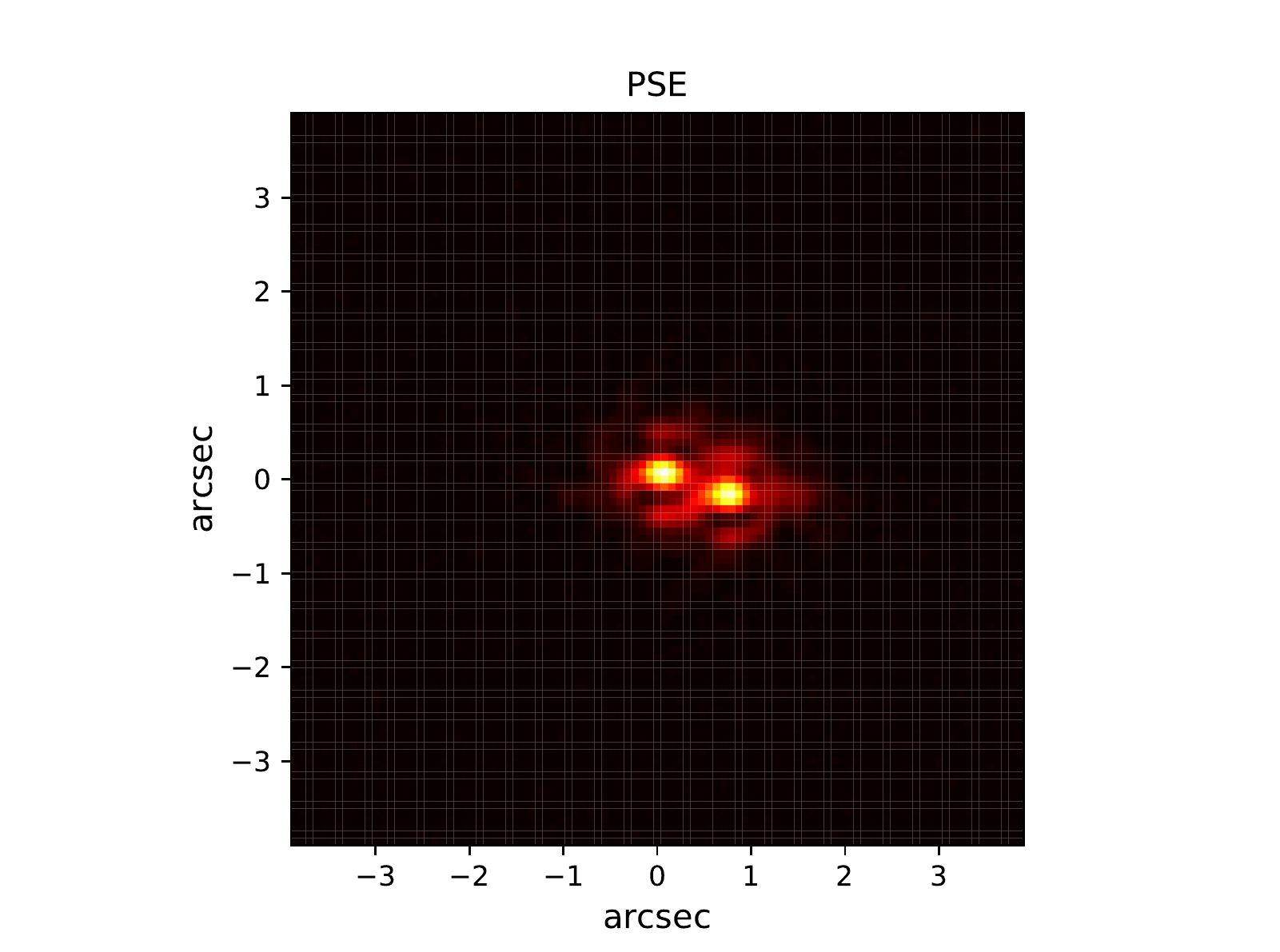} 
\includegraphics[width=65mm]{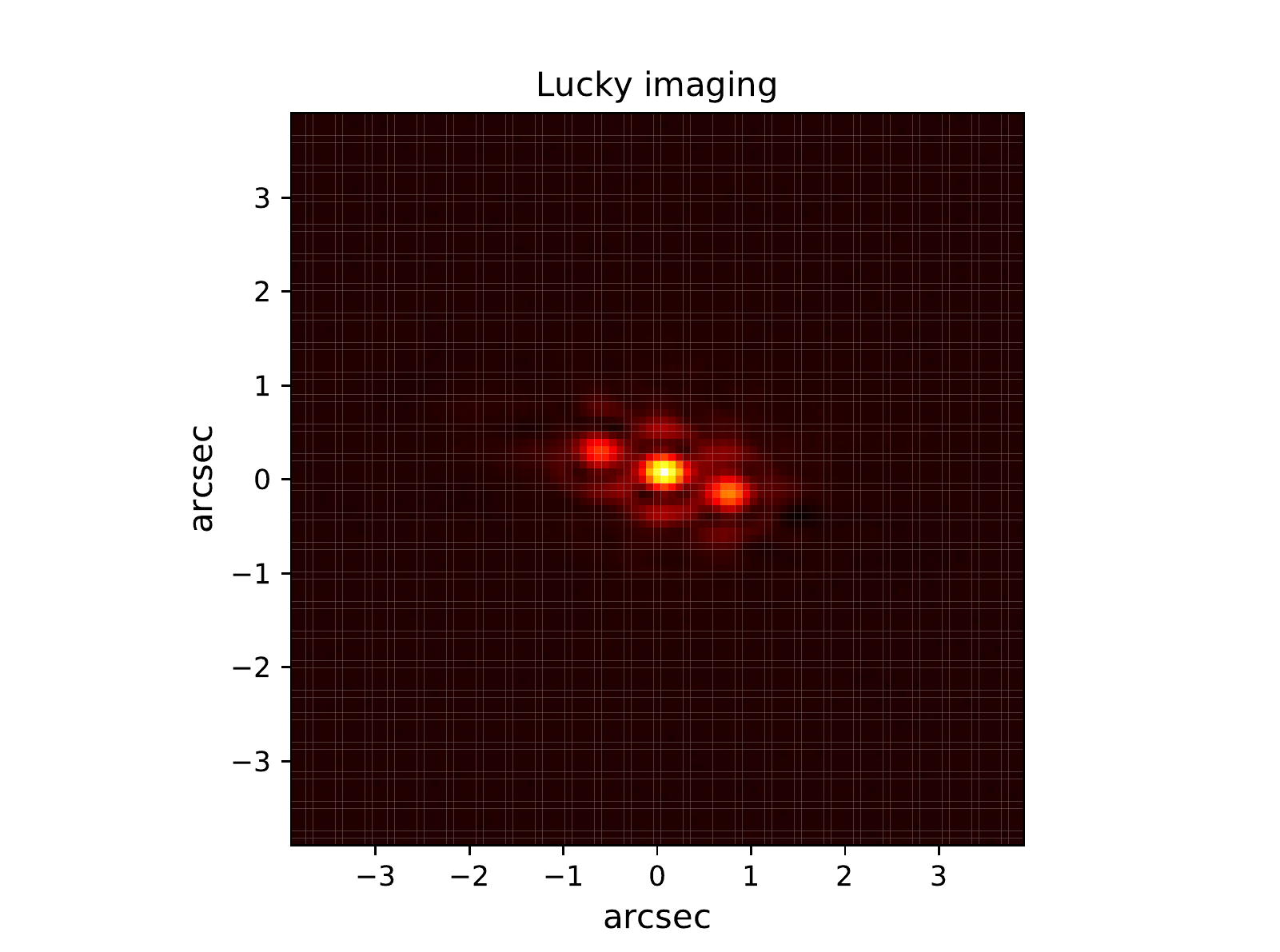} 
\end{center}
\caption{Reconstruction of the binary star i~Boo, having a small magnitude difference ($\Delta m=0.09$). Left: PSE reconstruction. Right: FL reconstruction. The latter shows a ghost companion due to misalignment of images in the shift-and-add process.}
\label{fig:fantom}
% script python : fantom.py
\end{figure}

The ``ghost'' effect is illustrated by the Fig.~\ref{fig:fantom}. It happens on images of binary stars having a small magnitude difference. In that case, the classical centroid calculations used in the methods of SA and LI are not efficient. The brightest pixel or speckle can correspond to one star or the other, depending on atmospheric fluctuations. Reconstructed images using this alignment show a ghost image at a position symmetric to the actual companion (Fig.~\ref{fig:fantom}, right), resulting into a quadrant ambiguity. To solve this problem, we align images by computing cross-spectra between consecutive images, as explained in Sect.~\ref{par:dataproc}. The cross-spectrum is a function which takes into account the complete structure of the images (not only the brightest speckle), and gives the global shift between them, providing that they are similar, i.e. that tip-tilt is the dominant effect. It works well if the time lag between images is short. The example on the double star i~Boo (Fig.~\ref{fig:fantom}, left) shows no ghost effect, and allows, in particular, to compute the magnitude difference between the two stars. In our target list, about 30\% of objects presented a ghost effect on the lucky-imaging image: all could be successfully reconstructed by the PSE technique.

\subsubsection*{Measurements on binary stars}
The pair separations $\rho$ and their position angle $\theta$ were measured on reconstructed images using the software GdPisco by J.L. Prieur (Observatoire Midi-Pyr\'en\'ees)\cite{gdpisco}. This software is used routinely to process speckle images from the PISCO instrument\cite{Scardia20} at the C2PU telescope. The magnitude difference was estimated by performing a fit of two Airy discs on the reconstructed images. Alternate values for the separation and position angle could be derived from this fit, and compared to GdPisco estimates. The agreement was found within the error bars.

Observations are presented in Table~\ref{table:mesur}. For each couple, we report the WDS number\cite{wdscatalog}, the discoverer designation and components involved (Col. 2), the star name (Col. 3),  the epoch of observation in Besselian years (Col. 4), the seeing measured by the CATS station\cite{Aristidi19} given at $\lambda=500$nm (Col. 5), the exposure time in ms (Col. 6), the total number of frames (Col. 7), the percentage of frames selected by the algorithm (Col. 8), the angular separation in arcsec with its uncertainty (Col. 9), the position angle (Col. 10) and the H magnitude difference (Col. 11). If an orbit is available, the orbit reference (same designation as in the WDS orbit catalog\cite{wdsorbitcat}) is given in Col. 12 as well as the computed separation and position angle (Cols. 13 and 14).

The magnitude difference could not be determined in a reliable way for the pairs WDS~16326+4007 and WDS~16581+1509, due to a strong astigmatism on images. For the case of WDS~17335+5734 the pair is too close to be seen on the reconstructed image. The separation and the position angle were derived from the modulus of the Fourier transform of the image (by adjusting a cosine function), and the absolute quadrant could not be determined. They are flagged with a asterisk in the Table~\ref{table:mesur}

\begin{sidewaystable}
{\scriptsize
\begin{tabular}{cccccccccccccc}
WDS & Disc. & Name & Epoch & Seeing & Exp  & Nb     & Select.   & $\rho$ & $\theta$  & $\Delta m$  & Orbit & $\rho_c$ & $\theta_c$ \\
    &  Id.  &      &       & ($''$) &(ms)  & frames & rate (\%) & ($''$) & ($^\circ$)&             &       & ($''$)   & ($^\circ$) \\ \hline
10200+1950 & STF1424AB & $\gamma$ Leo & 2016.461 &  1.5& 10  & 4000   & 20 &  4.70$\pm$0.02 & 126.7$\pm$0.4 & 2.2$\pm$0.1 & Pko2014& 4.716 & 126.3 \\
13007+5622 & BU1082 & 78 Uma       & 2016.477 &  1  & 30  & 14000  & 20 & 0.80$\pm$0.01 & 124.2$\pm$0.5 & 1.92$\pm$0.06 &  Dru2014 & 0.758	& 126.0 \\
13064+2109 & COU11AB & 39 Com       & 2016.477 &  1  & 50  & 14000  & 20 & 1.78$\pm$0.01 & 316.0$\pm$0.4 & 2.4$\pm$0.1 &  &  &  \\ 
13100+1732 & STF1728AB & $\alpha$ Com & 2016.458 &     & 15  & 13000  & 20 & 0.36$\pm$0.02 & 192.2$\pm$0.6  & 0.01$\pm$0.05 & Mut2015 & 0.373 &	192.2 \\ 
13198+4747 & HU644AB & ADS 8862     & 2016.472 & 0.7 & 20  & 10000  & 15 & 0.48$\pm$0.02 & 86.4$\pm$0.9 & 0.46$\pm$0.1 & WSI2015 &  0.534	& 82.2\\ 
13377+5043 & STF1770 & ADS 8979     & 2016.453 & 2.3 & 15  & 10000  & 4 & 1.61$\pm$0.01 & 121.0$\pm$0.5  & 3.4$\pm$0.1 &   &  &  \\ 
13377+5043 & STF1770 & ADS 8979     & 2016.453 &  2.3& 30  & 10000  & 4 & 1.67$\pm$0.01  & 120.7$\pm$0.6 & 3.3$\pm$0.1 &   &  &    \\ 
14411+1344 & STF1865AB & $\zeta$ Boo  & 2016.458 &     & 30  & 15000  & 20 & 0.36$\pm$0.02 & 290$\pm$1 & 0.13$\pm$0.06  & Sca2007&  0.384 & 288.5 \\ 
14417+0932 & STF1866 & ADS 9345     & 2016.477 & 0.8 & 50  & 8000   & 20 & 0.71$\pm$0.01 & 204.1$\pm$0.8  & 0.14$\pm$0.02 &   &  &   \\ 
15038+4739 & STF1909 & i Boo        & 2016.433 &     & 10  & 20000  & 5 & 0.73$\pm$0.01 & 72.5$\pm$0.4 & 0.09$\pm$0.03 & Izm2019 & 0.789 &	70.4 \\ 
15232+3017 & STF1937AB & $\eta$ Crb & 2016.450 &  & 15  & 13000  & 10 & 0.51$\pm$0.02  & 219$\pm$1 & 0.13$\pm$0.02 & Mut2010  & 0.569	& 218.9  \\  
15360+3948 & STT298AB & ADS 9716     & 2016.453 & 3   & 40  & 15000  & 10 & 1.12$\pm$0.01 & 184.2$\pm$0.9 & 0.36$\pm$0.02 & Izm2019 & 1.187	& 185.8\\  
16009+1316 & STT303AB & ADS 9880     & 2016.472 & 0.7 & 100 & 8000   & 20 & 1.61$\pm$0.01 & 173.7$\pm$0.4 & 1.54$\pm$0.02 & Izm2019 & 1.583 &	173.9\\  
16309+0159 & STF2055AB & $\lambda$ Oph & 2016.458& 0.7  & 30 & 13000  & 10 & 1.37$\pm$0.01 & 43.4$\pm$0.4 & 1.58$\pm$0.02 & Izm2019 & 1.396 &	42.3\\  
16326+4007 & STT313 & ADS 10111    & 2016.477 & 0.6 & 100 & 8000   & 20 & 0.93$\pm$0.01 & 130.2$\pm$0.4 &   &   &  &  \\  
16413+3136 & STF2084 & $\zeta$ Her  & 2016.450 &     & 10  & 10000  & 20 & 1.27$\pm$0.01 & 128.8$\pm$0.4 & 2.3$\pm$0.1  & Izm2019 & 1.283	& 125.4\\  
16581+1509 & STT319 & ADS 10277    & 2016.461 & 1   & 50  & 15000  & 4 & 0.84$\pm$0.01 & 2.2$\pm$0.04 &  &   &  &  \\ 
17053+5428 & STF2130AB & $\mu$ Dra    & 2016.461 & 1.5 & 50  & 12000  & 5 & 2.54$\pm$0.01  & 2.2$\pm$0.4 & 0.23$\pm$0.02 & Izm2019 &2.501 &	1.3\\ 
17237+3709 & STF2161AB & $\rho$ Her   & 2016.458 &     & 30  & 15000  & 20 & 4.10$\pm$0.02 & 321.1$\pm$0.4 & 1.7$\pm$0.1 &   &  &  \\  
17335+5734 & MLR571 & HR 6560      & 2016.477 &  0.7& 50  & 10000  & &  0.23$\pm$0.01$^*$ & 82$\pm$1$^*$ &    & Grf2013 &  0.089 & 112.5 \\  
17520+1520 & STT338AB & ADS 10850    & 2016.461 &  1.2& 30  & 10000  & 4 & 0.80$\pm$0.01 & 166.9$\pm$0.7 & 0.06$\pm$0.02 &  Pru2012 & 0.829 & 167.7 \\  
17571+4551 & HU235 & ADS 10934    & 2016.477 &  0.7& 50  & 10000  & 4 & 1.58$\pm$0.01 & 286.5$\pm$0.4 & 2.2$\pm$0.1 &    &  &   \\  
18025+4414 &BU1127AB& ADS 11010    & 2016.472 &  0.6& 100 & 8000   & 4 & 0.66$\pm$0.01 & 46.3$\pm$0.6 & 2.0$\pm$0.3 &Cve2016 & 0.701 & 45.4   \\  
18070+3034 & AC15AB & b Her        & 2016.454 &  2.5& 20  & 15000  & 20 & 1.32$\pm$0.02 & 326.7$\pm$0.5 & 2.54$\pm$0.02 & Jao2016 & 1.393 & 328.5 \\  
18101+1629 & STF2289 & ADS 11123    & 2016.461 & 1.5 & 30  & 13000  &  4 & 1.19$\pm$0.01 & 220.2$\pm$0.4 & 3.4$\pm$0.1 & Izm2019 &1.216& 218.6  \\  
18385+3503 & COU 1308 &    & 2016.477 & 0.7 & 100 & 10000  & 20 & 0.42$\pm$0.01 & 28.6$\pm$1.3  & 0.73$\pm$0.02 &   &   &   \\  
18570+3254 & BU648AB & ADS 11871    & 2016.454 & 2.5 & 20  & 13000  & 50 & 1.26$\pm$0.01 & 240.7$\pm$0.5 &1.8$\pm$0.2 & Izm2019 & 1.295 & 240.1   \\  
20020+2456 & STT395 & 16 Vul       & 2016.477 & 0.7 & 50  & 6000 & 20 & 0.75$\pm$0.01 & 127.0$\pm$0.4 & 0.32$\pm$0.03&  Zir2013 & 0.849 & 126.7 \\

\end{tabular}
}
\caption{Table of measurements.}
\label{table:mesur}
\end{sidewaystable}

\begin{figure}
\begin{center}
\includegraphics[width=80mm]{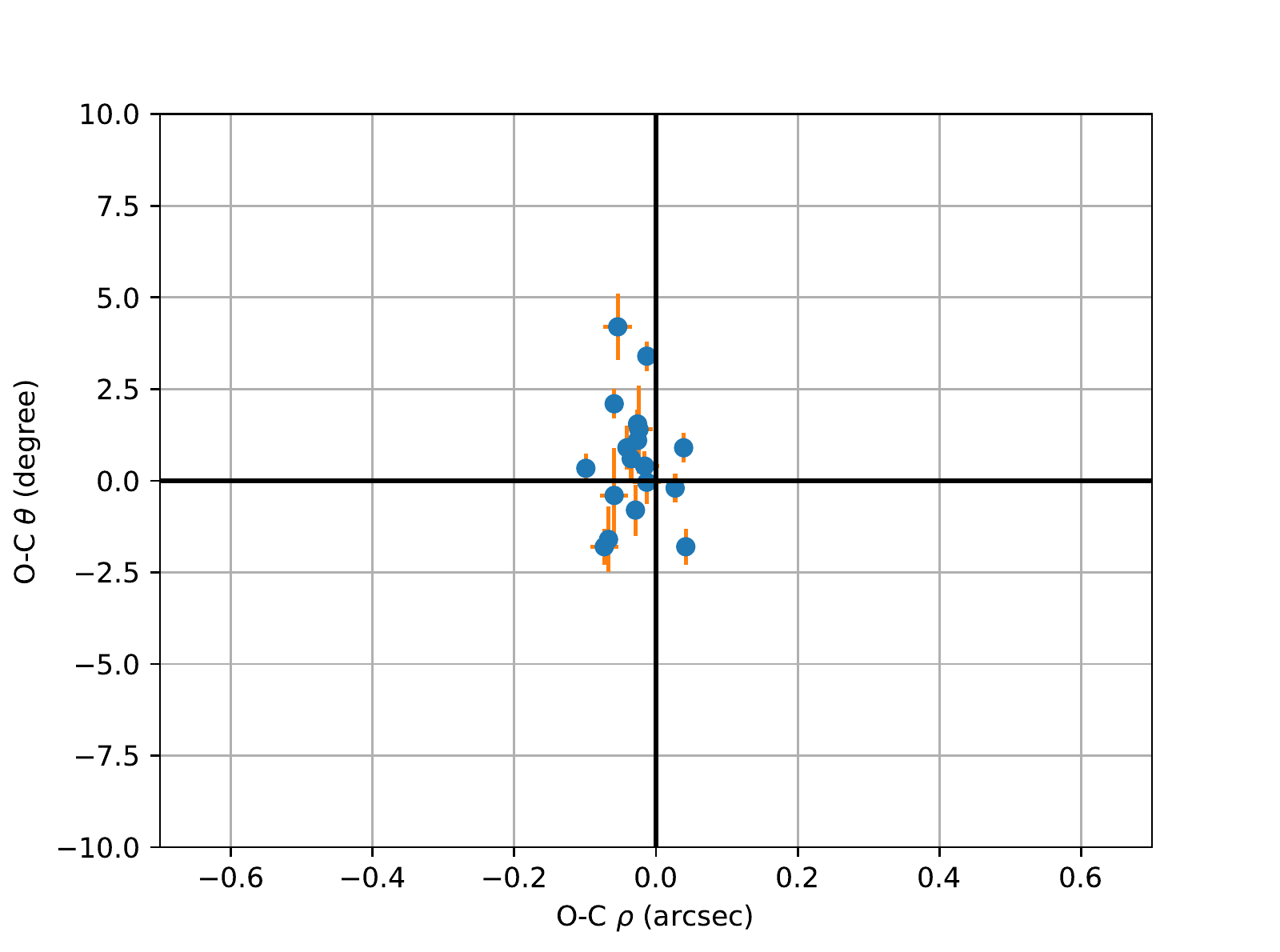}
\end{center}
\caption{Observed minus Computed (O--C) residuals of measurements of Table~\ref{table:mesur} with respect to available orbits. Error bars correspond to uncertainties given in cols 9--10 of Table~\ref{table:mesur}. The O-C for WDS~17335+5734, showing a large discrepancy for the position angle ($\Delta\theta=-30.5^\circ$) is not drawn on the graph.}
\label{fig:OC}
% script python : OCplot.py
\end{figure}

When an orbit is available (cols 12--14 of Table~\ref{table:mesur}), the Observed minus Computed (O--C) residuals $\Delta\rho$ and $\Delta\theta$ could be calculated. They are plotted in Fig.~\ref{fig:OC}. The median values computed with the residuals are $\Delta\rho_{\mbox{\scriptsize median}}=-0.02''$  and $\Delta\theta_{\mbox{\scriptsize median}}=0.36^\circ$. The small values obtained for these offsets provide a validation of our calibration procedure (see Sect.~\ref{par:dataproc}).

Large residuals have been found for the star WDS~17335+5734: $\Delta\rho=0.14''$, $\Delta\theta=-30.5^\circ$. The separation and position angle of this star were measured in the Fourier plane, as its separation is lower than the diffraction limit of the telescope. Looking at our data, we think that the orbit is not reliable, since it predicts a separation of $0.089''$ which is clearly too small (it would have been undetectable with our instrumentation).

%%%%%%%%%%%%%%%%%%%%%%%%%%%%%%%%%%%%%%%%%%%%%%%%%%%%%%%%%%%%%
\section{CONCLUSION}
\label{par:ccl}
In this paper, we have presented measurements of close binary stars in the infrared (H-band) obtained with a 1m telescope with a commercial near-IR camera. This was a first test run which allowed to check the limits of our experiment. The limiting magnitude will definitely increase when the CIAO bench will be fully functional, allowing larger exposure times. Observations of close binaries in the H-Band are difficult to find in the literature; in our target list of Table~\ref{table:mesur}, we found only two couples (HU~644AB and AC~15AB) whose relative photometry has been measured at this wavelength\cite{Henry93}. Though such observations are of importance for cool stars, especially close red dwarves with short orbital periods.

This presentation was also the first astronomical applications of our new lucky imaging technique, the PSE, which seem promising as it gives results comparable (and sometimes slightly better) than the Garrel FL algorithm, considered as one of the most efficient LI technique. Indeed the PSE has the advantage to be very fast: on a set of 6000~frames of size 128$\times$128, our Python-based PSE program takes $\sim$50s (with a selection rate of 20\%) while 470s are needed for the FL algorithm. This is a gain of CPU time of a factor $\sim$9, interesting for real-time processing during observations.

%%%%%%%%%%%%%%%%%%%%%%%%%%%%%%%%%%%%%%%%%%%%%%%%%%%%%%%%%%%%%
\section*{Acknowledgments}
We wish to thank the CIAO group, in particular F.X. Schmider, F. Vakili and F. Martinache for the time and energy they spent to write funding applications to buy the camera. This  research  has  made  use  of  the  Washington Double Star Catalog maintained at the U.S. Naval Observatory, and of the SIMBAD data-base, operated at CDS, Strasbourg, France. In particular, we thank Rachel Matson and Brian Mason, from the US naval observatory, for providing observation data for $\zeta$Uma. Thanks are also due to J.L. Prieur for providing his software Gdpisco that we used to reduce the data.
\bibliography{biblio}   %>>>> bibliography data in report.bib
\bibliographystyle{spiebib}   %>>>> makes bibtex use spiebib.bst

\end{document}